\documentclass[9pt,twocolumn,twoside,lineno]{article}

\usepackage{lipsum}
\usepackage{graphicx}
\usepackage{amsmath}
\usepackage{amsfonts}
\usepackage{amssymb}

\newcommand{\merge}{\textrm{merge}}
\newcommand{\all}{\textrm{all}}

\title{Resolution Limits for Detecting Community Change in Multilayer Networks}
\author{Michael Vaiana, Sarah Muldoon}

\begin{document}
\twocolumn[
  \begin{@twocolumnfalse}
    \maketitle
    \begin{abstract}
Multilayer networks capture pairwise relationships between the components of complex systems across multiple modes or scales of interactions.  An important meso-scale feature of these networks is measured though their community structure, which defines groups of strongly connected nodes that exist within and across network layers. Because interlayer edges can describe relationships between different modalities, scales, or time points, it is essential to understand how communities change and evolve across layers.  A popular method for detecting communities in multilayer networks consists of maximizing a quality function known as modularity. However, in the multilayer setting the modularity function depends on an interlayer coupling parameter, $\omega$, and how this parameter affects community detection is not well understood.  Here, we expose an upper bound for $\omega$ beyond which community changes across layers can not be detected. This upper bound has non-trivial, purely multilayer effects and acts as a resolution limit for detecting evolving communities.  Further, we establish an explicit and previously undiscovered relationship between the single layer resolution parameter, $\gamma$, and interlayer coupling parameter, $\omega,$ that provides new understanding of the modularity parameter space. Our findings not only represent new theoretical considerations but also have important practical implications for choosing interlayer coupling values when using multilayer networks to model real-world systems whose communities change across time or modality.

%
\end{abstract}
  \end{@twocolumnfalse}
]


\maketitle

Multilayer networks are quickly becoming the modeling framework of choice to represent complex interactions in large, multi-modal datasets.  A multilayer network is a rich generalization of a traditional network that captures interactions between nodes by separating each interaction type into its own layer together with interlayer coupling of nodes between layers \cite{Kivela2014,domenico2013mathematical,Boccaletti2014}.  Multilayer networks have found applications in a diverse range of settings such as neuroscience \cite{Muldoon2016,DeDomenico2017,BassetSporns2017,vaiana2017multilayer}, financial assets \cite{Bazzi2016,brummitt2015cascades}, congressional voting similarity \cite{Mucha2010,mucha2010communities}, social networks \cite{szell2010multirelational,bollen2011happiness,dickison2016multilayer}, and spreading processes \cite{cozzo2013contact,Sarzynska2014,DeDomenico2016}.

Recently there has been much interest in detecting communities in dynamic and multilayer settings \cite{Delvenne2010,bassett2013robust,Gauvin2014,domenico2015identifying,iacovacci2015mesoscopic,domenico2015structural,
matias2016statistical,ghasemian2016detectability,taylor2016enhanced,de2017community,peel2017ground,liu2018global}.
A community is a group of nodes with stronger connections to nodes within the group than to nodes external to the group, and the organization of the network into communities has strong implications for the function and structure of the system.  Communities in multilayer networks represent a balance between the community structure in and between layers, and detecting multilayer communities can provide insight into the network structure which is hidden at the level of the individual layers \cite{Mucha2010}.  Because multilayer communities can describe multiple interactions (throughout time, space, modality, etc.) between different layers of the network, it is especially important to understand how communities change and evolve across layers.

A popular class of algorithms attempt to optimize a quality function that measures how well a given partition of the network matches the underlying community structure.  Multiple quality functions have been developed from the perspective of network topology \cite{Newman2004,Newman2006}, information theory \cite{rosvall2008maps}, and statistical physics \cite{reichardt2006statistical,Lambiotte2014}, and the optimization of different quality functions can return different community partitions.  The first and most popular quality function is the modularity function \cite{Newman2004,Newman2006} which measures the number of internal community edges compared to a random network. However, Fortunato and Barth\'elemy exposed a fundamental problem with modularity maximization \cite{fortunato2007resolution} by showing that in single layer networks, there is a resolution limit such that communities that are small relative to the network can not be detected.  Later, Traag et al. \cite{traag2011narrow} showed that any method that relies on optimizing a global quality function suffers from a resolution limit, thereby demonstrating that the resolution limit represents a fundamental challenge for a large class of algorithms.  Other work relating statistical physics and modularity \cite{reichardt2006statistical,Lambiotte2014} introduced a tunable multiresolution parameter, $\gamma$, to the modularity function that can be used to control the resolution of community detection. In fact, it has been shown \cite{tibely2008equivalence,traag2011narrow} that several quality functions \cite{Newman2004,reichardt2006statistical,raghavan2007near,
arenas2008analysis,ronhovde2009multiresolution} can all be realized as a specific formulation of a generalized multiresolution modularity function, and this multiresolution modularity function has been widely adopted to mitigate the resolution limit problem.

Importantly, modularity maximization was one of the first methods to be extended to multilayer networks \cite{Mucha2010,bassett2013robust} through a simple modification of the multiresolution modularity function.  As such, it currently remains one of the most commonly used algorithms for performing community detection in the multilayer setting.  The multilayer modularity function includes two tunable parameters: the resolution parameter, $\gamma$, and an interlayer coupling parameter, $\omega,$ that controls the strength of the interlayer coupling, i.e. the edges that run between layers. The interlayer coupling allows communities to span across layers, and the balance between detecting community structure within and between layers is controlled by $\omega.$  When $\omega$ is small, the community structure of each layer will be preferred and nodes can easily switch communities between layers. When $\omega$ is large, nodes will prefer to stay in one community across layers and will be less compelled by the community structure within the layers.  While some work attempts to provide guidance on how to choose these parameters \cite{weir2017post,amelio2017revisiting}, little is known about how the choice of parameter values influences community detection.

Here, we expose a resolution limit on community detection in multilayer networks such that a change in community structure between two layers can not be detected. We show precisely how the interlayer coupling parameter, $\omega$, controls the ability to detect communities in multilayer networks and give an upper bound on $\omega$ beyond which it can be guaranteed that certain cross-layer community changes can not be detected. We demonstrate an explicit relationship between our bound and the previously established single layer resolution limit and show how our bound has non-trivial and purely multilayer effects.  Further, we show that $\omega$ is bounded above by a linear function in $\gamma$ establishing an explicit and previously unknown relationship between these parameters.

\section*{Multi-resolution Modularity in Single Layer Networks}
We first review modularity maximization in a traditional single layer network.  Given a network with an adjacency matrix, $A$, and a randomized version of the network, $R$, the (multi-resolution) modularity function is
\begin{equation}
\label{eq:multi-res-modularity}
Q(P) = \sum_{ij}(A_{ij} - \gamma R_{ij})\delta(i,j)
\end{equation}
where $P = \{M_1, M_2, \ldots, M_k\}$ is a partition of the network into communities, and $\delta(i,j)=1$ if $i$ and $j$ are in the same community and $0$ otherwise.  The intuition is that modularity is higher when nodes inside a community have stronger connections than random. The parameter, $\gamma$, controls the resolution of community detection.  When $\gamma$ is large, smaller communities are detected and when $\gamma$ is small, larger communities are detected.  Despite concerns related to the degeneracy of the function \cite{Good2010} and the ability of $\gamma$ to truly eliminate the effects of the resolution limit in networks with a heterogeneous distribution of community sizes \cite{lancichinetti2011limits}, it has been shown that modularity maximization offers a balance of speed and accuracy when tested against a variety of benchmark networks \cite{Yang2016}, and modularity maximization remains one of the most popular methods for performing community detection.

\section*{Community Structure in Multilayer Networks}
In order to discuss modularity maximization in the multilayer framework, we first give a brief introduction to multilayer networks. For a more comprehensive review see \cite{Kivela2014}. A multilayer network is a collection of vertices and edges separated into distinct layers.  We let $N = \{1,2, \ldots, n\}$ be the node set of the network, and we denote the layers by the greek letters $\alpha$ and $\beta.$  Then node $i$ on layer $\alpha$ is denoted as $i_{\alpha}.$  We let $w(i_{\alpha}, j_{\beta})$ denote the weight of the edge between $i_{\alpha}$ and $j_{\beta}.$  Such an edge is called an intralayer edge if $\alpha = \beta$ and an interlayer edge if $\alpha\neq \beta.$  We adopt the common assumption that interlayer coupling is \emph{diagonal} \cite{Kivela2014,  Bazzi2016}, that is, the only nonzero interlayer edges are those which connect a node to itself in another layer. The interlayer edges are then given by $B_{i\alpha\beta} = w(i_{\alpha},i_{\beta})$, which is the weight of the edge between node $i$ in layer $\alpha$ with itself in layer $\beta.$ Similarly, intralayer edges are given by $A_{ij\alpha} = w(i_{\alpha}, j_{\alpha}).$  Note that for a fixed $\alpha$, the matrix $A_{ij\alpha}$ is just the regular adjacency matrix for layer $\alpha.$

\begin{figure}
\centering
\includegraphics[width=.8\linewidth]{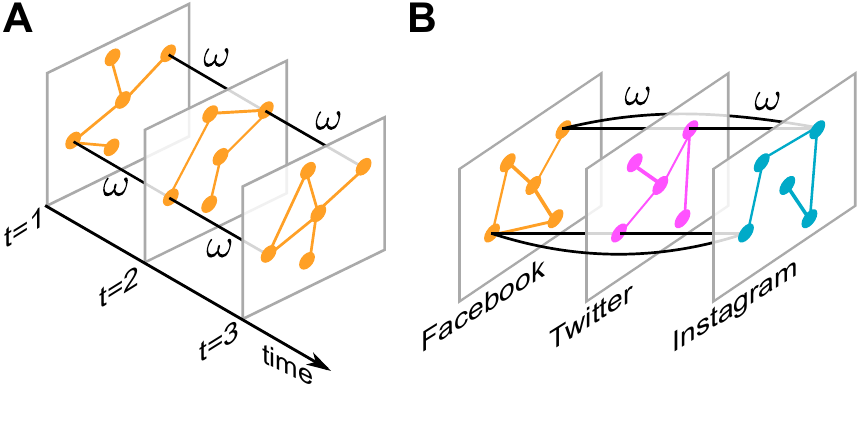}
\caption{Example multilayer networks. For clarity, interlayer coupling is only drawn for two nodes. \textbf{A})  A dynamic temporal network in which intralayer edges change over time. \textbf{B})  A multilayer network of online social interaction. Each layer represents a different mode of interaction.}
\label{fig:mln-examples}
\end{figure}

\subsection*{Modularity in Multilayer Networks}
A partition of a multilayer network is a partition of the vertices $i_{\alpha}$. Let $P= \{M_1, M_2, \ldots, M_k\}$ be such a partition, then the groups of vertices $M_r$ are call communities or modules. For a given multilayer network, the multilayer modularity function \cite{Mucha2010} measures how well a partition matches the communities of the network and is given by
\begin{equation}
\label{eq:multi-mod-full}
Q(P) = \sum_{ij\alpha} (A_{ij\alpha} - \gamma R_{ij\alpha})\delta({i_{\alpha}}, {j_{\alpha}}) +\sum_{i\alpha\beta}B_{i\alpha\beta} \delta({i_{\alpha}}, {i_{\beta}})
\end{equation}
where $R_{ij\alpha}$ is a null model of $A_{ij\alpha}, \gamma$ is a resolution parameter, and $\delta(i_{\alpha},j_{\beta})= 1$ if $i_{\alpha}$ and $j_{\beta}$ are placed in the same community in $P$ and 0 otherwise. We say that $\delta$ is the community function induced by partition $P$. It is important to note the community function is not the Kroneker delta function, although it is similar.

Often, for simplicity, it is assumed that the interlayer edges are constant \cite{weir2017post}, i.e. $B_{i\alpha\beta} = \omega$ for all $i,\alpha,\beta$ and for some scalar $\omega.$  We will assume this for the remainder of the article unless explicitly stated otherwise. In this case, the modularity function is given by
\begin{equation}
\label{eq:multi-mod}
Q(P) = \sum_{ij\alpha} (A_{ij\alpha} - \gamma R_{ij\alpha})\delta({i_{\alpha}}, {j_{\alpha}}) +\sum_{i\alpha\beta}\omega \delta({i_{\alpha}}, {i_{\beta}}).
\end{equation}
Under similar assumptions to those above, Bazzi et al. \cite{Bazzi2016} gave global bounds $\omega_{\textrm{min}} < \omega < \omega_{\textrm{max}}$ which represent extreme points in community detection. When $\omega > \omega_{\max},$ nodes never change communities across layers, and when $\omega< \omega_{min},$ the communities in each layer are determined entirely by the intralayer topology, and hence the multilayer aspect of the network is essentially ignored.  While useful, these global bounds say nothing about how well multilayer modularity detects community structure between any given pair of layers. Below we give a local bound on $\omega$ such that it can be guaranteed that modularity fails to detect community changes when $\omega$ is beyond this bound.

\subsection*{Changes in Community Structure Across Layers}
\begin{figure}
\centering
\includegraphics[width=.8\linewidth]{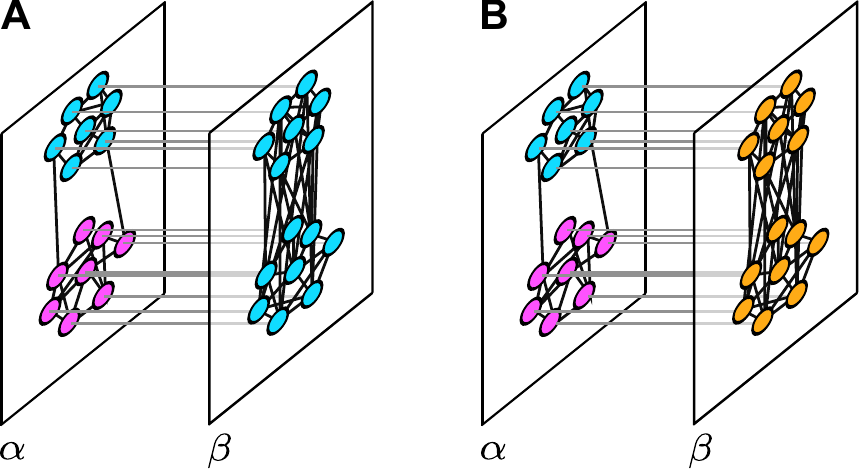}
\caption{Two examples of communities merging between layers ${\alpha}$ and ${\beta}$.  Color indicates the community assignment. We say a community survives a merger if its community assignment does not change across layers. \textbf{A})  Two communities in layer $\alpha$ merge in layer ${\beta}$. The blue community survives the merge.  \textbf{B})  Two communities in layer ${\alpha}$ merge together in layer ${\beta}.$  Neither of the two communities survive the merger.  }
\label{fig:merger}
\end{figure}

A fundamental goal of multilayer community detection is to understand how communities change between layers.  When a group of distinct communities in layer $\alpha$ merge into one single community in layer $\beta$, then there has been a change in community structure across layer $\alpha$ and $\beta.$  When interlayer edges are undirected, several communities merging into one community is equivalent to one community splitting into several smaller communities.  For this reason, a merger of communities represents the most basic type of community change that can be expected in a multilayer network.  We therefore focus on the behavior of multilayer modularity in the presence of community mergers. See Fig. \ref{fig:merger}. Note that we will assume that layer $\alpha$ and $\beta$ are connected by interlayer edges.

\subsection*{Interlayer Coupling Bounds}
We now give general bounds on the interlayer coupling parameter beyond which community mergers are undetectable by modularity. Let $K\subset N$ be a subset of the nodes of the network.  We are assuming that the nodes of $K$ form $t$ many communities in layer $\alpha$ and one large community in layer $\beta$. In other words, the nodes of $K$ merge between layer $\alpha$ and $\beta.$  Throughout, we will let $\{C_1,C_2, \ldots C_t\}$ denote the distinct communities of $K$ on layer $\alpha$ and we let $J$ be the remaining nodes of the network that are not in $K.$ 

\begin{figure}
\centering
\includegraphics[width=.8\linewidth]{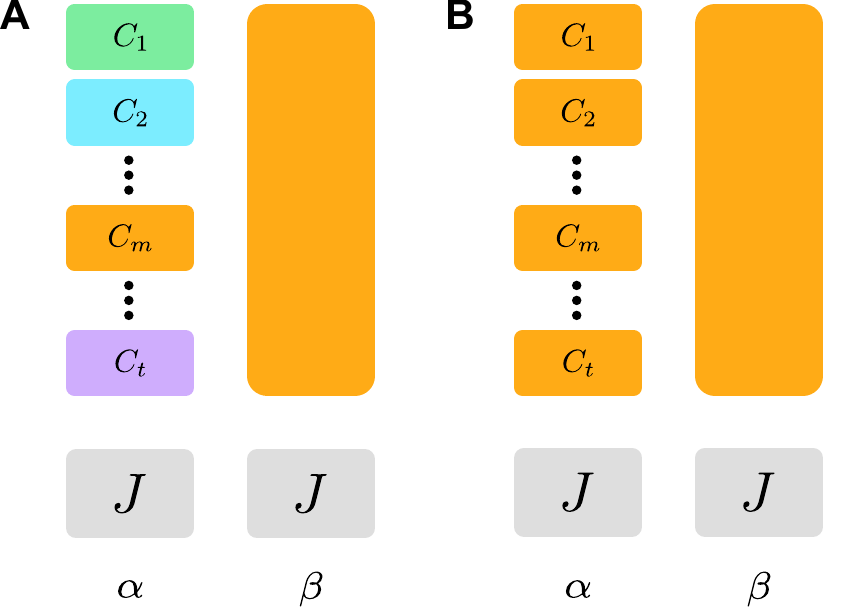}
\caption{Layers $\alpha$ and $\beta$ of a multilayer network in which the the nodes of $K$ form communities $C_1,\ldots C_t$ on layer $\alpha$ merge into a single community in layer $\beta.$  Color indicates community assignment.   \textbf{A}) The partition $P_{\merge}$ which correctly identifies the merger of the individual communities in layer $\alpha$ into one large community in layer $\beta.$  \textbf{B}) The partition $P_{\all}$ which erroneously groups all the nodes into one community in both layer $\alpha$ and $\beta$. Note each of the two partitions are arbitrary and identical on the remaining nodes $J.$ }
\label{fig:merge-vs-all}
\end{figure}

We would like to know if modularity can accurately identify the change in community structure of the nodes of $K$ between the two layers. Let $P_{\merge}$ be the partition which correctly identifies the merger of communities of $K$ and let $P_{\all}$ be the partition which places all the nodes of $K$ in one community in both layer $\alpha$ and $\beta.$ Importantly the two partitions are arbitrary but identical on the remaining nodes, $J$, and on all other layers of the network. See Fig. \ref{fig:merge-vs-all}.

Notice that $P_{\merge}$ is not unique since there are $t+1$ many ways a merge can happen: one of the $t$ many communities can survive the merge, or none can (see Fig. \ref{fig:merger}).  Locally, when considering only layers $\alpha$ and $\beta$, modularity is always higher when the largest community survives, so we assume $C_m$ is the largest community and that it survives. However, this assumption is not strictly necessary and only effects our results up to a constant. 

We now give an upper bound on the value of the interlayer coupling parameter beyond which community mergers are undetectable. In all computations for the remainder of the paper, the function $\delta$ will refer to the community function induced by the partition $P_{\merge}.$ The partitions agree everywhere except the $K$ nodes of layer $\alpha$ and are identical on layer $\beta.$ Thus from equation \ref{eq:multi-mod} we have
\begin{align}
Q(P_{\all}) &-Q(P_{\merge})
\\&=  \sum_{\substack{ij\in K \\ \delta(i_{\alpha},j_{\alpha}) = 0}} (A_{ij\alpha} - \gamma R_{ij\alpha}) + 2\sum_{\substack{i\in K\\ i\not\in C_{m}}} \omega
\\&= \sum_{\substack{ij\in K \\ \delta(i_{\alpha},j_{\alpha}) = 0}} (A_{ij\alpha} - \gamma R_{ij\alpha}) + 2\theta\omega
\label{eq:all-minus-merge}
\end{align}
where $\theta = |K| - |C_{m}|$ is the number of nodes of $K$ that are not in the largest community $C_{m}.$ Therefore $Q(P_{\merge}) > Q(P_{\all})$ if and only if
\begin{equation}
2\theta\omega < \sum_{\substack{ij\in K \\ \delta({i_{\alpha}}, {j_{\alpha}}) = 0}} (\gamma R_{ij\alpha} - A_{ij\alpha})
\label{eq:full-inequality}
\end{equation}
which gives
\begin{equation}
\label{eq:upperbound-all}
\omega < \sum_{\substack{ij\in K \\ \delta(i_{\alpha}, j_{\alpha}) = 0}} \frac{1}{2\theta}(\gamma R_{ij\alpha} - A_{ij\alpha}) \equiv \Omega.
\end{equation}
Thus modularity can detect the merger of the communities of $K$ if and only if $\omega < \Omega.$  Notice that equation \ref{eq:upperbound-all} is dependent only on layer $\alpha.$
Our immediate goal is to characterize this bound and understand its behaviors so we can determine which values of $\omega$  will allow us to detect community mergers.

\section*{Consequences of Interlayer Coupling on Modularity}
In the last section, we gave an upper bound on the parameter $\omega$ in terms of a generic null model $R_{ij\alpha}.$  We now specialize the upper bound, $\Omega,$ to the popular Newman-Girvan (NG) model \cite{Newman2004}.  We then show a certain type of equivalence between $\Omega$ and the single layer resolution limit discussed in \cite{fortunato2007resolution}.  Also, we show an explicit dependence of the coupling parameter, $\omega$, on the resolution parameter, $\gamma.$  Finally, we discuss the consequences of these results on multilayer modularity. 

Recall from equation \ref{eq:upperbound-all} that the upper bound, $\Omega$, is only dependent on the nodes of layer $\alpha.$  For convenience and clarity, we drop the subscript $\alpha$, but all quantities are implicitly computed with respect to this layer. Now, define $\kappa_{i}$ to be the degree of node $i$ (counting only intralayer edges) and $2m = \sum_{i}\kappa_{i}$ to be the degree of all nodes (on layer $\alpha$).
Then the NG model is defined by $R_{ij} = \frac{\kappa_{i}\kappa_{j}}{2m}.$

Recall that $\{C_1, C_2, \ldots C_t\}$ are the communities of the nodes of $K$ in layer $\alpha.$ Let $d_r$ be the degree of community $C_r,$ that is $d_r = \sum_{i\in C_r} \kappa_i.$  Let $e_r$ be the portion of the external degree of $C_r$ that connects to one of the remaining communities $C_s.$ 
%
%
%
%
%
With this notation in place, we compute
\begin{align}
\sum_{\substack{ij\in K \\ \delta({i}, {j}) = 0}} R_{ij} &= \gamma\left[ \sum_{ij\in K}\frac{\kappa_{i}\kappa_{j}}{2m} -\sum_{\substack{ij\in K\\ \delta({i}, {j}) = 1}}\frac{\kappa_{i}\kappa_{j}}{2m} \right]
\\&=\frac{\gamma}{2m}\left[\left(\sum_{r=1}^{t}d_{r}\right)^2  - \sum_{r=1}^{t} (d_{r})^2\right]
\\&=\frac{\gamma}{2m}\sum_{r\neq s}^{t}d_{r}d_{s}.
\label{eq:NG-bound}
\end{align}
Similarly,
\begin{align}
\sum_{\substack{ij\in K \\ \delta({i_{}}, {j_{}}) = 0}} A_{ij} \
&=\sum_{r=1}^t e_r.
\label{eq:A-bound}
\end{align}
Plugging equations \ref{eq:NG-bound} and \ref{eq:A-bound}  into equation \ref{eq:upperbound-all} gives that
\begin{equation}
\label{eq:omega_all}
\Omega = \frac{1}{2\theta}\left[\frac{\gamma}{2m_{}}\sum_{r\neq s}^{t}d_{r}d_{s} - \sum_{r=1}^t e_r\right].
\end{equation}
\subsection*{The Resolution Limit}
Recall that if $\omega > \Omega,$ then the community merge of the nodes of $K$ can not be detected by modularity.  Since $\omega$ is assumed to be non-negative when $\Omega < 0 < \omega$ we will not properly resolve the change in the communities of $K.$

We now show an explicit connection with the single layer resolution limit. Assume that $t=2$, that is, there are only 2 communities in $K.$  Define $l_r$ to be the internal degree of community $C_r.$  In \cite{fortunato2007resolution}  Fortunato et al. showed in the absence of the resolution parameter $\gamma$  that if
\begin{equation}
l_2 < \frac{2e_1m}{l_1d_1d_2}
\label{eq:single-layer-res-limit}
\end{equation}
then modularity (restricted only to layer $\alpha$) will be highest when the two communities of $K$ are grouped into one larger community.  Let $X = \frac{2e_1m}{l_1d_1d_2}$ be this limit and define $z = \frac{d_1d_2}{l_1l_2}.$ 
From equations \ref{eq:NG-bound}, \ref{eq:A-bound}, and \ref{eq:omega_all} we have

\begin{align}
\Omega &= \frac{1}{2\theta}\left[\frac{2d_1d_2\gamma}{2m} - e_1 - e_2\right]
\\&= \frac{1}{2\theta}\left[\frac{z \gamma l_1(l_2 -\frac{X}{\gamma} + \frac{X}{\gamma})}{m} - 2e_1 \right]
\\&=\frac{z l_1}{2\theta m}(\gamma l_2 - X)
\label{eq:Omega-explicit}
\end{align}
where $\frac{X}{\gamma}$ represents the single layer resolution limit of \cite{fortunato2007resolution} adjusted according to the parameter $\gamma.$ In the computations, we used the fact that $e_1 = e_2$ and $\frac{zl_1X}{m} = 2e_1.$

Notice that $\gamma l_2 -X <0$ precisely when $l_2 < \frac{X}{\gamma}$, that is precisely when the single layer resolution limit implies the two communities will be detected as a single large community.  Since all other quantities in equation \ref{eq:Omega-explicit} are positive, it follows that \begin{equation}
\label{eq:iff-gamma-l2-X}
\Omega > 0 \iff \gamma l_2 - X > 0
\end{equation}
This implies that the community merger can be detected by multilayer modularity if and only if the individual communities can be detected by single layer modularity.  However, this is a much stronger statement than it seems since the additional upper bound $\Omega$ places further constraints on multilayer modularity. In fact, \emph{there exists} a value of $\omega$ such that multilayer modularity can detect two merging communities if and only if single layer modularity can detect those communities.  Thus, multilayer modularity succeeds only if single layer modularity succeeds \emph{and} the appropriate value of $\omega$ is chosen.  Further, the addition of interlayer links and the existence of the upper bound $\Omega$ have purely multilayer effects which we now discuss.

\subsection*{Multilayer Effects of the Resolution Limit}
We draw four important conclusions from equations \ref{eq:omega_all}, \ref{eq:Omega-explicit}, and \ref{eq:iff-gamma-l2-X}. 
\begin{enumerate}
\item Multilayer modularity can resolve community mergers if and only if single layer modularity can resolve the individual communities.
\label{conclusion:iff}
\item The upper bound, $\Omega$, scales inversely with the degree of the layer from which it is computed and is linear in $\gamma.$ In particular, it is more difficult to detect the merger of communities that are small relative to their layer.
\label{conclusion:proportional}
\item There is an explicit dependence between the parameters $\gamma$ and $\omega$: community mergers can only be detected when $\omega < a\gamma + b$ where $a$ and $b$ are constants depending on the structure of the communities which merge and the degree of the corresponding layer. 
\label{conclusion:omega-gamma}
\item When the degree of the nodes of a communities that merge are small, so too is $\Omega.$  This is especially important in networks where nodes may have low or zero degree on some layers and have high degree on others.
\label{conclusion:zero-degree}
\end{enumerate}

Conclusion \ref{conclusion:iff} is not a simple restatement of the single layer resolution limit.  It implies that the ability to detect \emph{changes in community structure} between layers is constrained by the traditional single layer resolution limit.  This is particularly important for any multilayer network statistics that measure changes in community assignment across layers \cite{bassett2011dynamicreconfig,bassett2015learningautonomy}. Also, as previously discussed, even when single layer modularity can resolve the communities, the multilayer modularity function will detect the merger only if $\omega < \Omega.$   Conclusion \ref{conclusion:proportional} is of practical concern, as it gives guidance on when to expect modularity to fail to detect community changes or on what scale $\omega$ should be set to improve resolution for detecting changes.  Specifically, if the communities that merge are small compared to their layer, then $\omega$ also needs to be small in order to resolve the change.

As far as we are aware, conclusion \ref{conclusion:omega-gamma} is the first result of its kind which explicitly determines a relationship between the parameters $\omega$ and $\gamma.$  In particular, changes in community structure will not be detected unless $\omega < a\gamma + b.$  From equation \ref{eq:omega_all}, we see the intercept, $b$, is non-positive and proportional to the external degree of the merging communities.  The slope, $a$, is inversely proportional to total degree of the layer from which the communities merge and proportional to the pairwise product of the community degrees.  Having an explicit linear relationship between $\omega$ and $\gamma$ may drive future research in understanding the modularity landscape with respect to parameter choice \cite{weir2017post}.

Finally, conclusion \ref{conclusion:zero-degree} can have strong and non-trivial consequences for detecting changes in multilayer networks.  Real world multilayer networks have been shown to have many nodes with zero degree on some or many layers of the network \cite{Nicosia2015}.  This is not surprising, as each layer represents a type of interaction, and a node may have a preferred mode of interacting, giving it non-zero degree on one layer and zero degree on another.  However, a merger of a group of zero degree nodes on layer $\alpha$ to a community in layer $\beta$ will not be resolved by multilayer modularity since in this case the upper bound $\Omega$ will be 0.  This has important consequences for detecting changes in multilayer networks in which nodes may be inactive on one or more layers.  We make this concrete with an example by showing how the community structure in one layer can propagate to others, making it impossible for modularity to detect the changing structure.

\subsection*{Example 1}
\label{ex:example}
\begin{figure}[!ht]
\centering
\includegraphics[width=.8\linewidth]{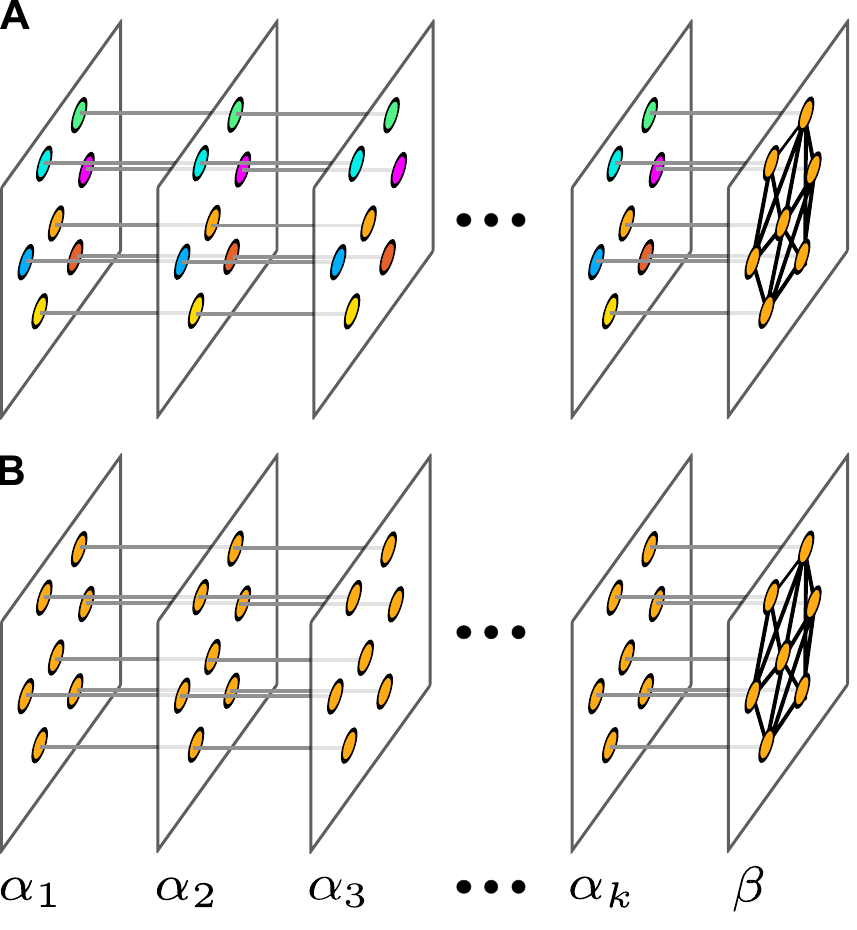}
\caption{A multilayer network with $k$ many layers, ${\alpha_i}$, with zero degree and one layer,  ${\beta}$, in which the nodes form a clique.  Color indicates community assignment in the given partition. \textbf{A}) The expected community structure based on the connectivity of the network. \textbf{B})  The result of community detection with any $\omega > 0.$}
\label{fig:example}
\end{figure}

Let $\mathcal{M}$ be a multilayer network with $k+1$ layers ${\alpha_1}, \ldots {\alpha_k}, {\beta}$ and $N$ nodes in each layer. Suppose that the vertices in layers ${\alpha_1}, \ldots {\alpha_k}$ all have 0 degree and that the vertices in layer $\beta$ form a clique, an all to all connected graph. Couple the network such that $B_{i\alpha_j\alpha_{j+1}} = \omega$ and $B_{i\alpha_k\beta} = \omega$ so only adjacent layers are connected. Then, any nonzero value of $\omega$ implies that $P_{\textrm{all}}$ has the maximum modularity amongst all the possible partitions of the network. Here $P_{\textrm{all}}$ is the partition which puts all nodes in all layers into one community. This can be seen by noting that all groupings of nodes in layer $\alpha_i$  result in no change in modularity since the nodes have $0$ degree.  On the other hand modularity is highest for layer $\beta$ when all the nodes are grouped into a single community. Finally, leaving a node in the same community across two layers contributes positively to modularity and it follows that modularity is highest when all nodes in all layers are assigned a single community. See Fig. \ref{fig:example}.

\subsection*{Example 2}
\label{ex:example2}
We now perform an explicit computation of the upper bound $\Omega$ on a toy network.  In this example, we assume the communities are disjoint cliques thus representing the strongest possible intralayer community structure.  The upper bound $\Omega$ in this case implies that modularity can not detect the merger of disjoint cliques when $\omega > \Omega.$  This has no single layer counter part in the sense that disjoint cliques do not suffer from a single layer resolution limit.

\begin{figure}[!ht]
\centering
\includegraphics[width=.8\linewidth]{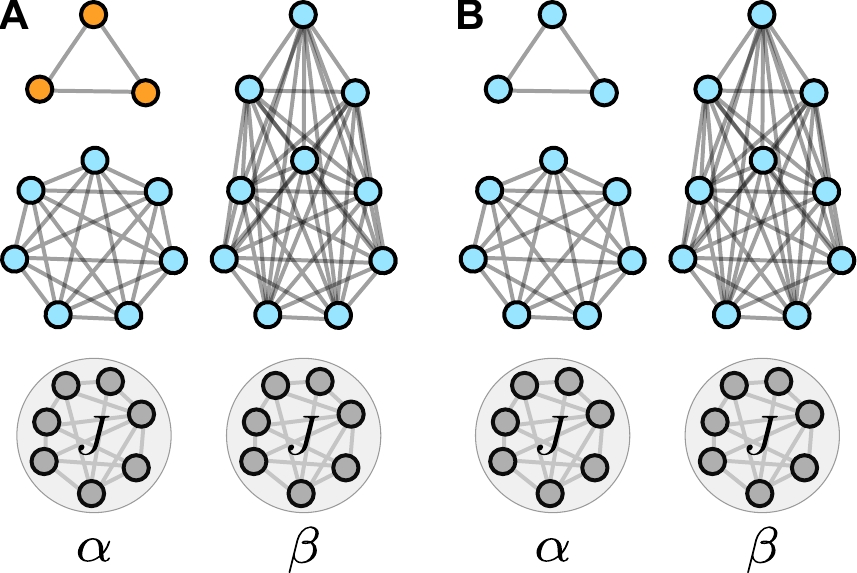}
\caption{A $n$ node multilayer network with $K=\{1,2,\ldots 10\}.$  The nodes of $K$ form disjoint cliques with 3 and 7 nodes in layer $\alpha$ and form a 10-clique in layer $\beta.$  Color indicates community assignment and the gray nodes represent the remaining nodes of the network. \textbf{A}) The partition $P_{\merge}$ which correctly identifies the communities of $K$ in both layer $\alpha$ and $\beta.$  \textbf{B}) The partition $P_{\all}$ which groups the nodes of $K$ into one community in both layers.}
\label{fig:toy-example}
\end{figure}

Consider a network with two layers ${\alpha}$ and ${\beta}$.  Let $K = \{1,2,\ldots, 10\}$ be the first 10 nodes of the network.  Assume in layer $\alpha$  the nodes of $K$ are partitioned into 2 disjoint cliques, a 3-clique and a 7-clique, and in layer 2 the nodes of $K$ form a 10-clique.  It is clear that the nodes of $K$ form 2 communities in layer 1 and merge into one large community in layer 2 (see Fig. \ref{fig:toy-example}).  Further assume that remaining nodes have arbitrary connectivity but total degree of $12.$ 
We now compute the upper bound, $\Omega,$ using equation \ref{eq:omega_all} as follows:
\begin{align}
\Omega &= \frac{\gamma}{2\theta}\frac{1}{2m_{}}\sum_{r\neq s}^2 d_{r}d_{s}
\\& = \frac{\gamma}{2\theta} \frac{2(6\cdot 42)}{48+12}
\\& =  \frac{\gamma}{14}\frac{504}{60} = 0.6\gamma
\end{align}


Here we find that, even in this simple toy model with strong intralayer community structure, in order to detect the change in communities, we must choose $\omega < 0.6 \gamma$.  This value is significantly less than global upper bound, $\omega_{\textrm{max}}$ given in \cite{Bazzi2016}, and importantly, smaller than one's intuition might expect given the strong intralayer community structure.



\section*{Discussion}
As multilayer networks continue to emerge as the predominate tool to model complex multi-modal, multi-scale relationships, it is increasingly important to understand the assumptions and limitations of such models.  The most popular method for detecting communities in multilayer networks is modularity maximization, but the modularity function depends on two parameters that must be selected by the user:  the resolution parameter, $\gamma$, and the interlayer coupling parameter, $\omega$.  

Here, we have demonstrated that there is an exact upper bound, $\Omega$, such that a merge of two communities is only detectable by modularity when $\omega < \Omega.$  Further, we have shown that this upper bound is dependent upon the choice of resolution parameter $\gamma$, resulting in a linear relationship between the two parameters: $\omega < a\gamma + b$.  These findings have very practical implications for those building multilayer networks from experimental data.  For example, in cases where the interlayer coupling values are measured experimentally, it might be necessary to introduce a global scaling of such values to ensure that for a given resolution value, the values of $\omega$ are within the proper bounds to detect cross layer changes in community structure.

While these findings introduce a relationship between $\gamma$ and $\omega$ and provide practical guidance for choosing parameters, this work only considers the case of a single community merger across two layers of the network.  In real-world multilayer networks, one expects multiple changes in community structure across multiple layers.  In order to ensure the best detection of community changes, one would therefore need to optimize the choice of $\gamma$ and over all layers and either choose the value(s) of $\omega$ accordingly between each pair of layers, or select a global $\omega$ that is within the bounds for all pairs of linked layers.

The selection of the resolution and coupling parameters for performing multilayer modularity maximization must be chosen with care in order to properly detect dynamic and inter-modal changes in community structure in multilayer networks. We present important bounds for these parameters and encourage those building multilayer networks from experimental data to carefully take these bounds into consideration when interpreting the results of dynamic community detection in such models.


\bibliographystyle{plain}
\bibliography{main}

\end{document}